\documentclass[preprint,amsmath,amssymb]{revtex4}
\newcommand{\me}{\mathrm{e}}

\newcommand{\dif}{\mathrm{d}}
\usepackage{pstricks}
\usepackage{color}
\usepackage{graphicx}
\usepackage{amsmath}

\newcommand\eqn[1]{(\ref{#1})}      
\newcommand\Eqn[1]{Eq.~(\ref{#1})}  
\newcommand{\bm}[1]{\mbox{\boldmath${#1}$}}
\renewcommand{\me}{\mathrm{e}}

\renewcommand{\dif}{\mathrm{d}}

\begin{document}
\title{Thermal Conductivity and Sound Attenuation in Dilute Atomic
Fermi Gases}

\author{Matt Braby, Jingyi Chao and Thomas Sch\"afer}
\address{Physics Department \\
North Carolina State University \\
Raleigh, NC 27695, USA}

\date{\today}
\begin{titlepage}
\renewcommand{\thepage}{}          

\begin{abstract}
We compute the thermal conductivity and sound attenuation length
of a dilute atomic Fermi gas in the framework of kinetic theory.
Above the critical temperature for superfluidity, $T_{c}$, the 
quasi-particles are fermions, whereas below $T_c$, the dominant 
excitations are phonons. We calculate the thermal conductivity 
in both cases. We find that at unitarity the thermal conductivity 
$\kappa$ in the normal phase scales as $\kappa \propto T^{3/2}$.
In the superfluid phase we find $\kappa \propto T^{2}$. At high 
temperature the Prandtl number, the ratio of the momentum and 
thermal diffusion constants, is 2/3. The ratio increases as the 
temperature is lowered. As a consequence we expect sound attenuation
in the normal phase just above $T_c$ to be dominated by shear viscosity. 
We comment on the possibility of extracting the shear viscosity 
of the dilute Fermi gas at unitarity using measurements of the 
sound absorption length. 

\end{abstract}

\maketitle
\end{titlepage}

\section{Introduction}
\label{intro}

 Cold, dilute Fermi gases in which the interaction between the atoms 
can be tuned using an external magnetic field provide a new paradigm 
for strongly correlated quantum fluids \cite{Bloch:2007,Giorgini:2008}. 
Recently, there has been significant interest in the transport properties 
of these systems \cite{Schafer:2009dj}. This interest was triggered by 
three independent developments. The first was the experimental 
observation of almost ideal hydrodynamic flow in the normal phase 
of a cold Fermi gas in the limit of infinite scattering length 
\cite{oHara:2002}. The second was the discovery of almost ideal flow 
in a completely different system, the quark gluon plasma created in 
heavy ion collisions at the Relativistic Heavy Ion Collider (RHIC) 
\cite{Adler:2003kt,Back:2004mh,Adams:2004bi}. Both of these experiments 
constrain the shear viscosity of the fluid. Remarkably, while the 
absolute value of the shear viscosity of the two systems differs by 
about 26 orders of magnitude, the ratio of shear viscosity $\eta$ 
to entropy density $s$ is very similar, $\eta/s\simeq (0.1-0.5)\,
\hbar/k_B$, where $\hbar$ is Planck's constant and $k_B$ is Boltzmann's 
constant \cite{Schafer:2007pr,Turlapov:2007,Schaefer:2009px,Dusling:2007gi,Romatschke:2007mq}.

 The third development was the theoretical discovery of a new 
method to compute the shear viscosity of strongly coupled field 
theories \cite{Policastro:2001yc,Son:2007vk}. This method is based
on the holographic duality between four dimensional field theory
and string theory in higher dimensional spaces \cite{Maldacena:1997re}.
Using the holographic equivalence one can show that in a large class 
of field theories the strong coupling limit of $\eta/s$ is equal to 
$\hbar/(4\pi k_B)\simeq 0.08\,\hbar/k_B$. This value is remarkably 
close to the experimental results for $\eta/s$ quoted above. Initially 
the value $\hbar/(4\pi k_B)$ was derived only for scale invariant 
relativistic field theories, but it was later shown that the same 
value of $\eta/s$ is obtained in the strong coupling limit of 
many non-conformal and non-relativistic field theories 
\cite{Son:2008ye,Balasubramanian:2008dm}. 

 In this work we will use kinetic theory to compute another important 
transport property of a dilute atomic Fermi gas, the thermal conductivity 
$\kappa$. We will consider both the high temperature as well as the 
low temperature, superfluid, phase. Our results complement earlier 
studies of the shear and bulk viscosity in the high and low temperature 
phase. The shear viscosity was studied in \cite{Bruun:2005,Rupak:2007vp}. 
In the unitarity limit the bulk viscosity of the high temperature phase 
vanishes \cite{Son:2005tj}. In the superfluid phase there are three bulk 
viscosities, one of which is non-vanishing in the unitarity limit. The 
bulk viscosities in the superfluid phase were recently computed in 
\cite{Escobedo:2009bh}. Combining our results for the thermal 
conductivity with the existing calculations of the shear viscosity we 
can study the relative size of momentum diffusion and thermal diffusion. 
The ratio of these quantities is known as the Prandtl number 
\begin{equation}
\label{Pr}
{\it Pr} = \frac{c_{P}\eta}{\rho\kappa} \, , 
\end{equation}
where $c_P$ is the specific heat at constant pressure and $\rho$ 
is the fluid mass density.

 The Prandtl number also controls the relative importance of shear 
viscosity and thermal conductivity in sound attenuation. Measurements 
of sound attenuation have the potential to provide new experimental 
constraints on transport properties of dilute atomic gases. Current 
estimates of the shear viscosity are based on the study of scaling flows, 
such as collective modes or the expansion out of a deformed trap 
\cite{Schaefer:2009px}. These experiments do not directly constrain 
the density independent part of the shear viscosity. This is a 
problem, because kinetic theory predicts that the shear viscosity 
at high temperature is only a function of temperature, and not of 
density. Sound attenuation, on the other hand, is sensitive to this 
contribution. Using our results for the Prandtl number it is possible 
to use measurements of sound attenuation to constrain the shear viscosity. 
Conversely, in the regime where shear viscosity is well constrained
it is possible to extract thermal conductivity from sound attenuation. 
First sound was observed in \cite{Joseph:2006}, and ideas for 
detecting second sound are discussed in \cite{Taylor:2009}.

 This paper is organized as follows. In Section \ref{sec:exc}, we will 
review the quasi-particle interactions in dilute atomic Fermi gases, and
in Section \ref{sec:kappa}  we outline the calculation of thermal 
conductivity using kinetic theory. In Sections \ref{sec:prandtl} and
\ref{sec:sound_att} we compute the Prandtl number and sound attenuation
length. We end with a few comments on the possibility of extracting 
the sound attenuation length from experiments with optical traps
in Section \ref{sec:concl}. Appendix \ref{app:trial} contains 
supplementary material regarding the trial functions used for solving 
the linearized Boltzmann equation, and in Appendix \ref{app:vir}
we collect some thermodynamic relations.

\section{Elementary Excitations}
\label{sec:exc}

\begin{figure}[t]
\begin{center}
\includegraphics{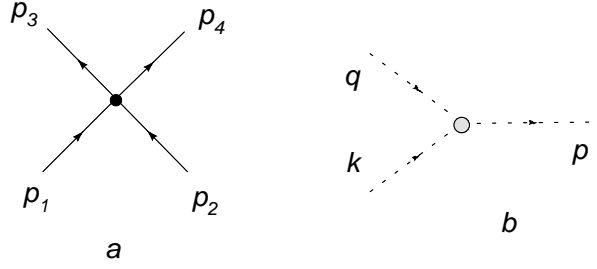}
\caption{\label{fey}
This figure shows the main processes that determine the thermal 
conductivity in high and low temperature phase, respectively. 
Fig.~(a) shows the local four-fermion interaction, and Fig.~(b) 
shows the three-phonon interaction.}
\end{center}
\end{figure}

 Kinetic theory is based on the existence of well defined quasi-particle
excitations. In the dilute Fermi gas quasi-particles exist both at 
very high and at very low temperature. In the high temperature limit
the relevant degrees of freedom are spin 1/2 non-relativistic atoms 
with mass $m$. In the dilute limit the scattering cross section 
is dominated by elastic $s$-wave scattering. The scattering amplitude
can be described in terms of a delta-function interaction $V(x-x')=
C_0\delta(x-x')$, see Fig.~\ref{fey}. Here, $C_0=4\pi a/m$ is  
related to the scattering length. The $s$-wave cross section is 
determined by sum of all two-body diagrams. The cross section 
is 
\begin{equation}
\label{sigmaf}
  \sigma = \int\dif\Omega\,
  \frac{a^{2}}{1+a^{2}q^{2}}\, , 
\end{equation}
where $\mathbf{p}_{1},\mathbf{p}_{2}$ and $\mathbf{p}_{3},\mathbf{p}_{4}$ 
are the ingoing and outgoing momenta and $\mathbf{q}$ is the relative 
momentum $\mathbf{q}=(\mathbf{p}_{1}-\mathbf{p}_{2})/2$. The relative 
momentum of the outgoing particles is $\mathbf{q}'=(\mathbf{p}_{3}-
\mathbf{p}_{4})/2$ with $|\mathbf{q}'|=|\mathbf{q}|$. The solid angle 
is defined in terms of the angle between the relative momenta 
$\mathbf{q}$ and $\mathbf{q}'$, $d\Omega=d\cos(\theta_{\mathbf{q}
\mathbf{q}'})d\phi_{\mathbf{q}\mathbf{q}'}$.

The characteristic temperature of the system is the Fermi 
temperature $T_F=k_F^2/(2m)$, where $k_F$ is the Fermi momentum 
defined by $k_F^3=3\pi^2 n_{tot}$ and $n_{tot}$ is the density 
of both spin states. The strength of the interaction can be 
characterized by the dimensionless parameter $(k_Fa)$. The regime 
of small and negative $(k_Fa)$ is the BCS limit. In this limit 
the Fermi gas becomes a BCS superfluid at a critical temperature 
$T_c$ which is exponentially small compared to $T_F$. The 
ratio $T_c/T_F$ increases with $|k_Fa|$. In the unitary limit 
$|k_Fa|\to\infty$ the critical temperature is $T_c\simeq 0.15T_F$
\cite{Burovski:2006}. The regime of positive $(k_Fa)$ is called 
the BEC limit. For small and positive $(k_Fa)$ the system is 
composed of weakly interacting bosons with mass $2m$ and density 
$n_{tot}/2$. Superfluidity occurs at the Einstein temperature 
$T_c\simeq 0.21 T_F$. 

 In the superfluid phase the low energy excitations are 
phonons, Goldstone modes associated with the spontaneous 
breaking of the $U(1)$ phase symmetry. At unitarity the 
effective lagrangian for the phonon field is strongly
constrained by scale invariance. Son and Wingate showed
that \cite{Son:2005rv},
\begin{eqnarray}
\label{phi_lag}
\mathcal{L}_{\phi}&=&\frac{1}{2}
  (\partial_{0}\phi)^{2}-\frac{1}{2}v_{s}^{2}(\bm{\nabla}\phi)^{2}
   -g_{3}\big[(\partial_{0}\phi)^{3}
   -9v_{s}^{2}\partial_{0}\phi(\bm{\nabla}\phi)^{2}\big] \\ \nonumber
&& -\frac{3}{2}g_{3}^{2}\big[(\partial_{0}\phi)^{4}
   + 18v_{s}^{2}(\partial_{0}\phi)^{2}(\bm{\nabla}\phi)^{2}
   - 27v_{s}^{4}(\bm{\nabla}\phi)^{4}\big]+ \ldots \, , 
\end{eqnarray}
where $g_{3}=\pi v^{3/2}\xi^{3/4}/(3^{1/4}8\mu^{2})$, $\mu$ 
is the chemical potential and $v_s$ is the speed of sound given by
$v_s^{2}=2\mu/(3m)$. At unitarity $\mu=\xi E_F$ with $\xi\simeq
0.44$ \cite{Carlson:2003}. At low momentum the phonon dispersion 
relation is linear, $\varepsilon_{p}=v_sp$. We will see that the
thermal conductivity is sensitive to non-linearities in
the dispersion relation. We will write the dispersion relation 
as $\varepsilon_{p}=v_sp(1+\gamma p^{2})$, where the parameter 
$\gamma$ can be expressed in terms of the coefficients of certain 
higher order terms in the lagrangian of Son and Wingate, 
$\gamma = -\pi^2 \sqrt{2\xi} (c_1 + 1.5 c_2)/k_F^2$. Rupak and 
Sch\"afer computed $c_1$ and $c_2$ using an epsilon expansion
\cite{Rupak:2008xq}. They find $c_1 \simeq -\sqrt{2}/(30 \pi^2 
\xi^{3/2})$ and $c_2\simeq  0$, corresponding to $\gamma = 
1/(30 m \mu)$. Other estimates of $\gamma$ can be found in 
\cite{Schakel:2009,Zwerger:2009,Salasnich:2008}.

 We will study the contribution to thermal conductivity from 
the phonon splitting process $\varphi\rightleftharpoons\varphi +
\varphi$, see Fig.~\ref{fey}. For $\gamma>0$ this process is 
kinematically allowed. The matrix element for the three phonon 
process is 
\begin{equation}
  i\mathcal{M}_{3}=
  g_{3}\Big(p_{0}[Q\cdot K]+k_{0}[Q\cdot P]+q_{0}[P\cdot K]\Big) \, , 
\end{equation}
where we have defined the four vector product $Q\cdot K=q_{0}
k_{0}-9v_s^{2}\bm{q}\cdot \bm{k}$. If $\gamma$ is negative then 
the dominant contribution arises from $\varphi+\varphi
\rightleftharpoons\varphi + \varphi$. We recently studied this 
process in a relativistic superfluid, the color-flavor-locked 
phase of QCD \cite{Braby:2009dw}. We found that the $2\rightleftharpoons 
2$ processes are controlled by $s,t,u$-channel exchanges with almost 
on-shell intermediate state phonons. This means that the dominant
process can be viewed as $1\leftrightharpoons 2$ phonon splitting 
followed by $2\leftrightharpoons 1$ phonon absorption. 

\section{Thermal Conductivity}
\label{sec:kappa}

 Thermal conductivity is defined by the relation between the heat 
current $\bm{h}$ and the temperature gradient, $\bm{h}=-\kappa\bm{\nabla}T$. 
In kinetic theory the heat current is expressed in terms of the 
quasi-particle distribution function,
\begin{equation}
\label{ht}
\bm{h}=\nu\int \dif\Gamma\,\varepsilon_p \bm{v}_p\delta f_p
\end{equation}
where $\nu$ is the degeneracy factor, $\dif \Gamma = \dif^3 p/(2\pi)^3$,
$\bm{v}_p$ is the particle velocity and $\delta f$ is the deviation
of the distribution function from its equilibrium value, $\delta f = 
f-f^0$, with $f_0(p) = 1/(\me^{(\varepsilon_p - \mu)/T} \pm 1)$ for
bosons and fermions, respectively. Note that in the case of phonons
there is no conserved particle number and $\mu = 0$. The correction 
to the distribution function is linear in $\bm{\nabla}T$ and we
can define 
\begin{equation}
\label{df}
 \delta f_p = -\frac{f_0(p)(1\pm f_0(p))}{T^2}\bm{\chi}\cdot \bm{\nabla}T\, . 
\end{equation}
With this form we can rewrite \Eqn{ht} as
\begin{equation}
\label{ht2}
 \bm{h} = - \frac{\nu \bm{\nabla} T}{3T^2} \int \dif \Gamma\, 
      \varepsilon_p\, \bm{v}_p\cdot \bm{\chi} f_0(p)(1\pm f_0(p))
\equiv -T \bm{\nabla}T 
    \langle \varepsilon_p\, \bm{v}_p| \bm{\chi}\rangle \, ,
\end{equation}
where we have defined an inner product as
\begin{equation}
\label{inner}
\langle \bm{a}|\bm{b} \rangle = \frac{\nu}{3T^3}\int \dif \Gamma\, f_0(p)
(1\pm f_0(p))\bm{a} \cdot \bm{b} \, . 
\end{equation}
We can then write the thermal conductivity as
\begin{equation}
\label{kappa}
\kappa = T\langle \varepsilon_p\, \bm{v}_p| \bm{\chi}\rangle \, . 
\end{equation}
There are certain constraints on the the form of $\bm{\chi}$.
The Boltzmann equation conserves energy, momentum, and, in the 
case of a conserved charge, particle number. A non-trivial constraint
on $\bm{\chi}$ arises due to momentum conservation. We have 
\begin{equation}
\label{constraint}
0 = \nu \int \dif \Gamma\, \bm{p}\, \delta f_p =
-\frac{\nu}{T^2}\int \dif \Gamma\, \bm{p} 
 (\bm{\chi}\cdot  \bm{\nabla}T)  f_0(p)(1\pm f_0(p))
= -T \bm{\nabla T}\langle \bm{p}|\bm{\chi}\rangle \, ,
\end{equation}
which implies that $\langle \bm{p}|\bm{\chi}\rangle = 0$. As we can 
see from the constraint and \Eqn{kappa}, any term in $\varepsilon_p 
\bm{v}_p$ that is linear in $\bm{p}$ will not contribute to the 
thermal conductivity. In particular, if the kinetic description
is governed by a quasi-particle with exactly linear dispersion 
relation then the thermal conductivity vanishes. 

 The function $\bm{\chi}$ is determined by solving the Boltzmann equation
\begin{equation}
\label{Boltzmanna}
  \frac{\dif f}{\dif t}=\frac{\partial f}{\partial t}
       +\bm{v}_{p}\cdot \bm{\nabla} f=C[f] \, , 
\end{equation}
where $C[f]$ is the collision integral. Schematically we can write  
\begin{equation}
 C[f] = \int \dif \Gamma_{n-1} \,
 w({\it in};{\it out})
D_{{\it in}\leftrightarrow{\it out}}\, , 
\end{equation}
where $\dif \Gamma_{n-1}$ is the $(n-1)$ particle phase space (all
the particles not labeled by momentum $p$), $w$ is the transition 
probability, and $D_{\rm{in\leftrightarrow out}}$ contains the distribution 
functions.
Taking the convective derivative on the left-hand side of the Boltzmann
equation and focusing only on terms proportional to the temperature 
gradient we can write
\begin{equation}
\frac{\dif f}{\dif t}
   = \frac{f_0(1\pm f_0)}{3T} \bm{\alpha}_p\cdot \bm{\nabla} T
 \equiv \frac{f_0(1\pm f_0)}{3T}|\bm{\alpha}_p \rangle 
  \cdot \bm{\nabla}T
\end{equation}
where $\bm{\alpha_p}$ will depend on equilibrium properties of the 
quasi-particle system. In the case of either particles with linear dispersion
(like phonons) or particles with non-relativistic dispersions and a chemical
potential, we can write,
\cite{Braby:2009dw},
\begin{equation}
\bm{\alpha_p}  \sim \frac{\varepsilon_p \bm{v_p}}{T} + A{\bm p}
\end{equation}
where $A$ depends only on the thermodynamics of the system.  We can 
linearize the right-hand side of \Eqn{Boltzmanna} in $\bm{\chi}$ as
\begin{equation}
 C[f] = C[f_0] + C[\delta f] = C[\delta f] 
 = \int \dif \Gamma\, w({\it in};{\it out}) 
             \delta D_{{\it in}\leftrightarrow {\it out}}
 \equiv \frac{f_0(1\pm f_0)}{3T}
         {\hat C}|\bm{\chi}\rangle \cdot \bm{\nabla}T \, , 
\end{equation}
where we have defined the linearized collision operator, $\hat{C}$, 
acting on the state $|\bm{\chi}\rangle$. We will specify $\hat{C}$ in more 
detail below. We can now write the linearized Boltzmann equation as
\begin{equation}
  \label{boltz2}
|\bm{\alpha}_p\rangle = {\hat C}|\bm{\chi}\rangle \, .
\end{equation}
The thermal conductivity is now determined by solving \Eqn{boltz2}
for $|\bm{\chi}\rangle$ and then computing $\kappa$ using 
\Eqn{kappa}. In practice, we will adopt a variational procedure. 
Using the constraint we can show that $T \langle \bm{\chi}|\bm{\alpha}_p
\rangle = \langle\, \varepsilon_p \bm{v}_p|\bm{\chi} \rangle$ for
both fermions and bosons.  The thermal conductivity can then be 
written in two alternative ways 
\begin{equation}
\label{kappa2}
 \kappa = T\langle\, \varepsilon_p \bm{v}_p|\bm{\chi} \rangle 
\hspace{0.5cm}{\rm and}\hspace{0.5cm}
\kappa  =  T^2\langle \bm{\chi}|\hat{C}|\bm{\chi}\rangle \, , 
\end{equation}
where we have used the Boltzmann equation, \Eqn{boltz2}, to derive the 
second equality. These two representations form the basis of the variational 
principle. Consider a trial state $|\bm{g}\rangle$. The Schwarz 
inequality implies that 
\begin{equation}
 \langle \bm{g}|\hat{C}|\bm{g}\rangle 
 \langle \bm{\chi}|\hat{C}|\bm{\chi}\rangle 
   \geq |\langle \bm{g}|\hat{C}|\bm{\chi}\rangle|^2 \, . 
\end{equation}
Using \Eqn{kappa2} and \Eqn{boltz2}, we can write this as
\begin{equation}
\label{kvar}
\kappa \geq T^2\frac{|\langle \bm{g}|\bm{\alpha}_p\rangle|^2}
   {\langle \bm{g}|\hat{C}|\bm{g} \rangle}
= \frac{|\langle \bm{g}|\varepsilon_p \bm{v}_p \rangle|^2}
   {\langle \bm{g}|\hat{C}|\bm{g} \rangle} \, . 
\end{equation}
Using \Eqn{kappa2} we observe that an exact solution of 
the linearized Boltzmann equation saturates the inequality. 
We can then choose to expand the trial function $\bm{g}$ in 
orthogonal polynomials and take advantage of that fact that 
$\bm{g}$ is a dimensionless vector-valued function to write
\begin{equation}
\label{poly}
\bm{g} = \bm{x}\sum_{s\neq0}^N b_s^{(F,H)} B_s^{(F,H)}(x^2) \, 
\end{equation}
 where $\bm{x} = \bm{p}/\Delta$, $x=|\bm{x}|$, and $\Delta$ is some 
characteristic  energy scale in the problem (e.g. $\Delta =T/v_s$ for 
massless particles and $\Delta = \sqrt{2mT}$ for massive particles). 
The polynomials are such that $B_0 = 1, B_1 = x^2 + a_{10}$, 
$B_2 = x^4 + a_{21}x^2 + a_{20}$, etc., and
$(F,H)$ labels the basis functions in the fermion and phonon case.  
The orthogonality condition is 
\begin{equation}
A^{(F,H)}_{s} \delta_{st} =  \int \dif x 
   f_0(x)(1\pm f_0(x)) x^4 B^{(F,H)}_s(x^2) B^{(F,H)}_t(x^2)\, . 
\label{ortho}
\end{equation}
The type of quasi-particle involved enters via the statistical 
factor $(1\pm f_0(x))$ and through the dispersion relation in 
$f_0(x)$. The specific forms of $B^{(F,H)}_m(x^2)$ are given 
in Appendix \ref{app:trial}. Note that we have enforced the momentum 
constraint by excluding the $s=0$ term in \Eqn{poly}.

\subsection{Normal Phase}

 In this section we will specify the matrix elements and the 
collision operator in the case of atom-atom scattering in the 
normal phase of the dilute Fermi gas. We will scale all the 
momenta as $x_i = p_i/\sqrt{2mT}$, use the Fermi-Dirac distribution 
functions and set $\nu=2$. The numerator of \Eqn{kvar} is 
\begin{eqnarray}
\langle \bm{g}|\varepsilon_p \bm{v}_p \rangle
 &=& \frac{2}{3T^3}\int \dif \Gamma\,  f_0(p)(1-f_0(p)) 
    \varepsilon_p \bm{v}_p \cdot \bm{g} \\ \nonumber
 &=& \frac{4m}{3\pi^2}\sum_{s\neq0} b_s^F \int \dif x\,  
  f_0(x)(1-f_0(x)) x^6 B_s^F(x^2)\\ \nonumber
&=& \frac{4m}{3\pi^2}\sum_{s\neq0}b_s^F A_{s1}^F \delta_{s1} =
 \frac{4m}{3\pi^2}b_1^F\,A_{11}^F
\end{eqnarray}
where we have used the orthogonality condition. The matrix element 
of the collision operator is 
\begin{equation}
\langle \bm{g}|{\hat C}|\bm{g} \rangle =
\frac{2}{3T^4}\sum_{s,t\neq0}b^F_{s}b^F_{t}\int
  \frac{\dif^{3}P\dif^{3}q}{(2\pi)^{6}}\, \dif \Omega\, 
  |\bm{v}_1-\bm{v}_2|\, \frac{a^{2}}{1+a^{2}q^{2}}\, 
  D^0_{2\leftrightarrow2} \, 
   \bm{g}_1 \cdot \left[\bm{g}_1 +\bm{g}_2 - \bm{g}_3 - \bm{g}_4\right]\, . 
\end{equation}
Here we have written the transition probability in terms of the 
cross section \Eqn{sigmaf} and the flux factor $|\bm{v}_1-\bm{v}_2|$. 
The integration variables are the pair momentum $\bm{P}=\bm{p}_1+
\bm{p}_2$, the relative momentum $\bm{q}=(\bm{p}_1-\bm{p}_2)/2$, and
the solid angle $\Omega$ is defined below \Eqn{sigmaf}. We have also 
defined  $D^0_{2\leftrightarrow 2} =f_0(x_1)f_0(x_2)[1-f_0(x_3)]
[1-f_0(x_4)]$. We can write 
\begin{figure}[t]
\begin{center}
\includegraphics[width=8.8cm]{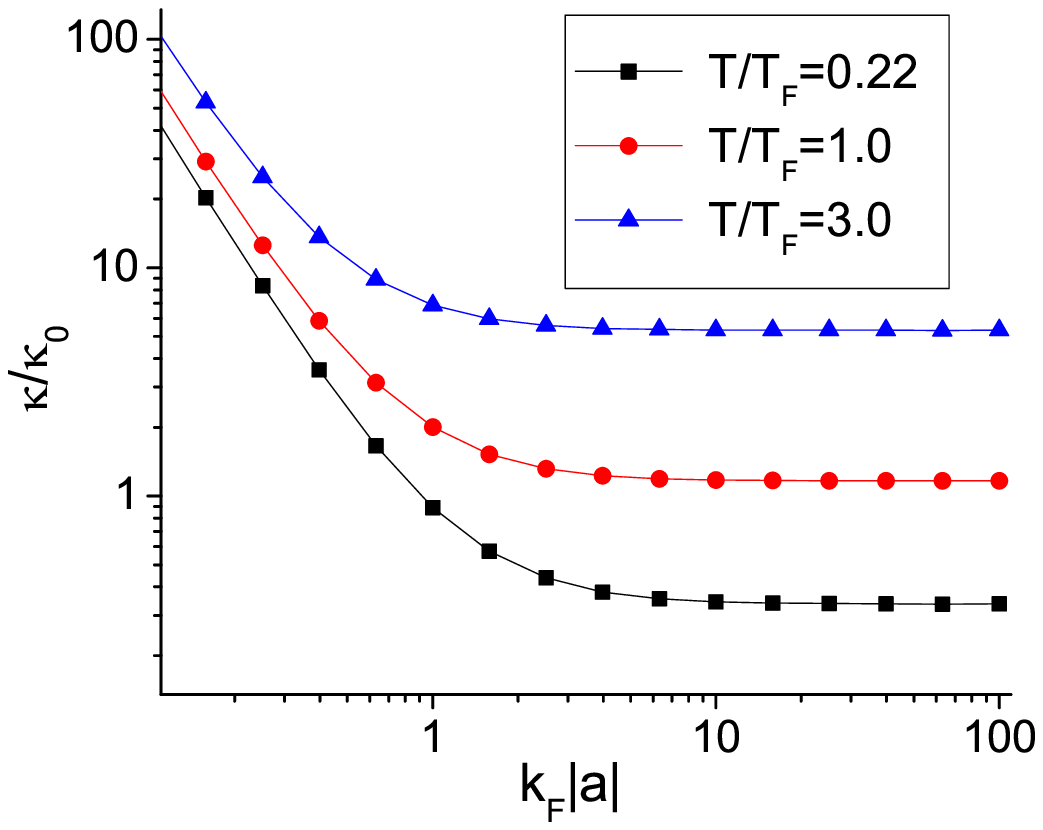}
\includegraphics[width=8.8cm]{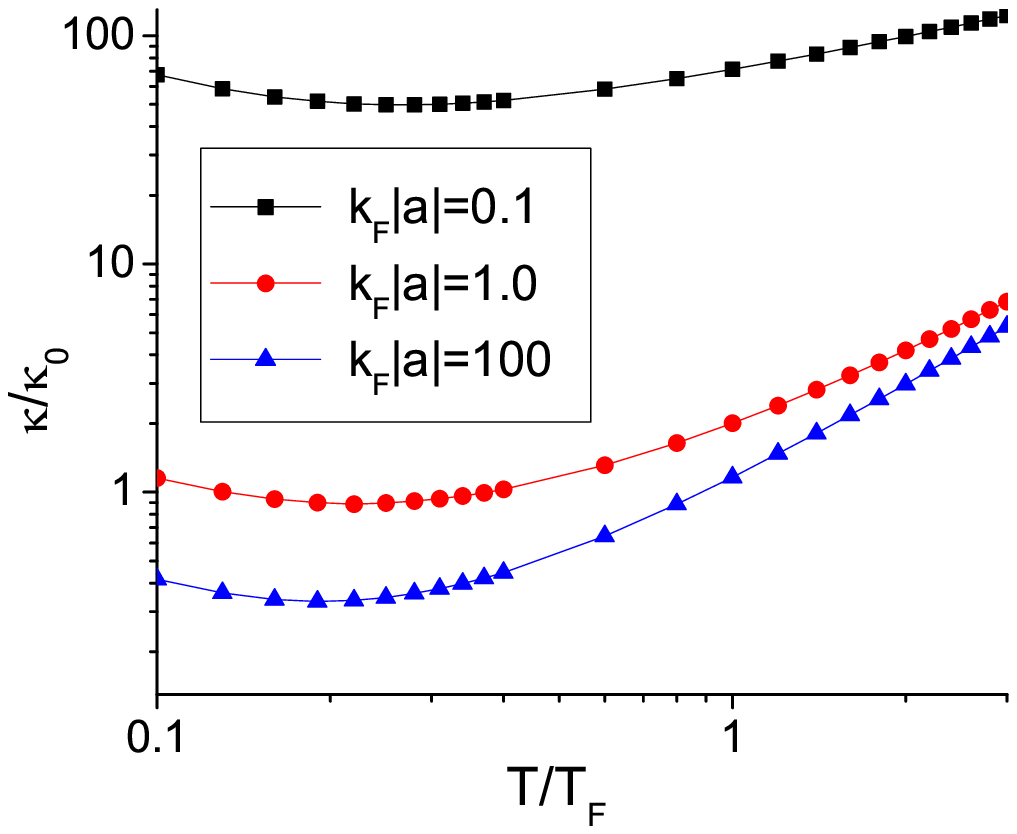}
\end{center}
\caption{Thermal conductivity in the normal phase of a dilute 
Fermi gas in units of $\kappa_0=m^{1/2}T_F^{3/2}$. In the left 
panel we show $\kappa/\kappa_{0}$ as a function of $T/T_F$ for 
different values of $k_Fa$. In the right panel we show $\kappa/
\kappa_{0}$ as a function of $k_Fa$ for different values of 
$T/T_F$. 
\label{kappa_fermi}}
\end{figure}
\begin{eqnarray}
\label{collintN}
\langle \bm{g}|{\hat C}|\bm{g} \rangle 
&=&\frac{2^{11/2}m^{3/2}}{3T^{3/2}}
\sum_{s,t\neq0}b^F_{s}b^F_{t}\int
  \frac{\dif^{3}x\dif^{3}y}{(2\pi)^{6}}\, \dif \Omega \,
  \frac{mTa^{2}y}{1 + 2mTa^{2} y^2}
  \, D^0_{2\leftrightarrow2}\, 
  \bm{B}^{F}_s\cdot\bm{B}^{F}_t  \\ 
&=& \frac{2^{11/2}m^{3/2}}{3T^{3/2}}
\sum_{s,t\neq0}b^F_{s}b^F_{t}M_{st}^F \, , \nonumber 
\end{eqnarray}
where $x = |\bm{P}|/\sqrt{2mT}$, $y = |{\bm q}|/\sqrt{2mT}$ and 
$\bm{B}^{F}_{t}=\big[B^F_{t}(x_{1})\bm{x}_1 + B^F_t(x_2)\bm{x}_2
- B^F_t(x_3)\bm{x}_3 - B^F_t(x_4)\bm{x}_4\big]/2$ . We finally get 
\begin{equation}
\kappa_F \geq \frac{\langle \bm{g}|\varepsilon_p \bm{v}_p \rangle^{2}}
    {\langle \bm{g}|{\hat C}|\bm{g} \rangle}
  =\frac{\kappa_0}{3\cdot2^{3/2}\pi^{4}}
   \left(\frac{T}{T_F}\right)^{\frac{3}{2}}D_F\left(k_Fa,T/T_F\right)\, , 
\end{equation}
where we have defined $\kappa_0 = m^{1/2}T_F^{3/2}$ and
\begin{equation}
D_F\left(k_Fa,T/T_F\right) \equiv 
   \frac{(b^F_1)^2 (A_{11}^F)^2}{\sum_{s,t\neq0} b_s^F b_t^F M_{st}^F}\, . 
\end{equation}
We can now consider a finite basis and optimize the inequality with 
respect to the coefficients $b_s^F$. We obtain $D_F\left(k_Fa,T/T_F\right)
=(A_{11}^F)^2 (M^{F})^{-1}_{11}$ where $(M^F)^{-1}_{11}$ refers to the 
(1,1)-component of the matrix inverse of $M$. Note that this result is 
equivalent to using the matrix form of the Boltzmann equation in 
a finite basis. The function $D_F$ must typically be calculated 
numerically. We can analytically evaluate $D_F$ in the limit of 
high temperature $T/T_F \gg 1$ and by keeping only the leading 
term in the polynomial expansion. We get 
\begin{equation}
\kappa_{F}=\begin{cases}
    \frac{75}{128\sqrt{\pi}}\sqrt{\frac{T}{m}}\frac{1}{a^{2}} 
   & \ \ \ |a|\rightarrow0 \\
    \frac{225}{128\sqrt{\pi}}m^{\frac{1}{2}}T^{\frac{3}{2}}  
   & \ \ \ |a|\rightarrow\infty
         \end{cases}
\label{kappa_highT}
\end{equation}
We have tried to improve this result using higher order terms
in the polynomial expansion, but the correction is very small, 
$\delta\kappa/\kappa<2\%$. This is consistent with the results 
in \cite{Bruun:2006}, where the authors observed very rapid convergence 
for the shear viscosity in the high temperature phase. Numerical 
results are shown in Fig.~\ref{kappa_fermi}. Note that the thermal 
conductivity depends only on the square of the scattering length
so we plot the thermal conductivity as a function of $k_F|a|$. We 
observe that, qualitatively, the temperature dependence of $\kappa$
is the same for all values of $k_F|a|$. In the unitarity limit 
$\kappa$ grows as $\sim T^{3/2}$ for $T\gg T_F$. In the BCS 
limit the temperature dependence is $\kappa\sim T^{1/2}$. The 
thermal conductivity has a minimum at $T/T_F \simeq (0.2-0.3)$, 
and increases for very small temperatures. This rise is related 
to Pauli blocking. We should note, however, that except in the 
BCS limit the results in this section are not reliable for $T\ll 
T_F$. The right panel shows that at fixed temperature $\kappa$ 
decreases as the strength of the interaction increases, and that 
$\kappa$ becomes insensitive to $k_Fa$ as we approach the 
unitarity limit $k_Fa\to\infty$.

\subsection{Superfluid Phase}

\begin{figure}[t]
\begin{center}
\includegraphics[width=10cm]{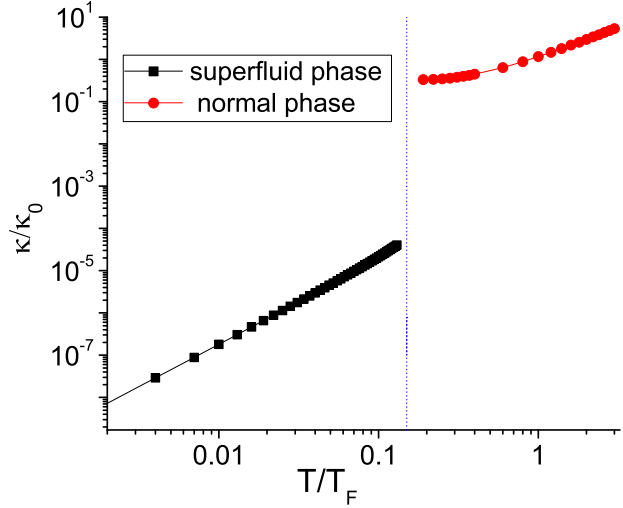}
\caption{Thermal conductivity $\kappa/\kappa_0$ of the dilute Fermi
gas at unitarity as a function of $T/T_{F}$. We show the calculations
in the high and low temperature limit, extrapolated to $T\simeq T_c$. 
The position of the critical temperature is indicated by the dashed 
line. 
\label{kappa_both}}
\end{center}
\end{figure}

 In this section we study the thermal conductivity in the superfluid 
phase. We will concentrate on the unitarity limit $k_Fa\to\infty$. In 
the BCS limit the critical temperature is exponentially small, and
the thermal conductivity is dominated by fermionic quasi-particles
even in the case $T<T_F$. In the extreme BEC limit the thermal 
conductivity reduces to that of an almost ideal Bose gas 
\cite{Kirkpatrick:1985}. Following the procedure outlined in the 
case the normal phase, we can write the numerator of \Eqn{kvar} as
\begin{eqnarray}
\langle \bm{g}|\varepsilon_p \bm{v}_p \rangle 
  &=&  \frac{1}{3T^3}\int \dif \Gamma f_0(p)(1+f_0(p))
        \varepsilon_p\, \bm{v}_p \cdot \bm{g}\\ \nonumber
  &=& \frac{2\gamma}{3\pi^{2}}\frac{T^{3}}{v_s^{4}} 
         \sum_{s\neq 0} b_s^H \int \dif x 
              f_0(x)(1+f_0(x)) x^6 B_s^H(x^2) \\ \nonumber
 &=& \frac{2\gamma}{3\pi^{2}}\frac{T^{3}}{v_s^{4}}b_1^H A_{11}^H\, ,
\end{eqnarray}
where we have used $\nu=1$, scaled the momenta as $x = vp/T$ and kept only 
the leading order terms in the 
parameter $\gamma$ that governs non-linearities in the dispersion 
relation. We note that the result is indeed proportional to $\gamma$, 
and $\kappa$ vanishes in the case of an exactly linear dispersion
relation as discussed before. The matrix element of the collision operator is 
\begin{figure}[t]
\begin{center}
\includegraphics[width=9cm]{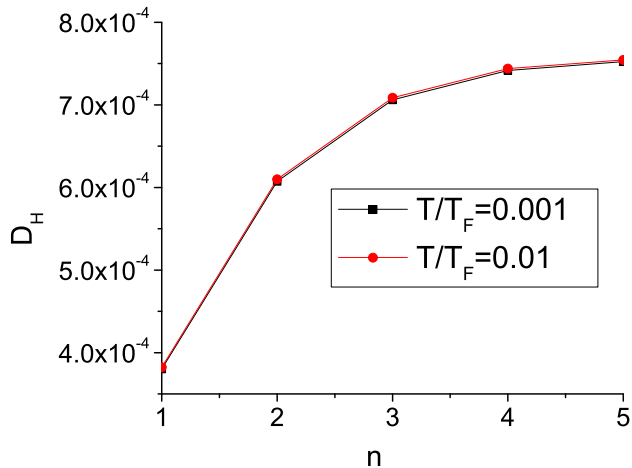}
\end{center}
\caption{
Variational function $D_H$ as a function of the number of 
terms in the polynomial expansion of $\tilde{g}(x)$. We 
show $D_H$ for two different temperatures.  
\label{kappa_bose}}
\end{figure}
\begin{eqnarray}
\langle \bm{g}|\hat{C}|\bm{g}\rangle
 &=& \frac{1}{3T^4}\sum_{s,t\neq 0} b^H_s b^H_t 
   \int \dif \Gamma_{pkq}\, 
      |{\cal M}_3(p,k,q)|^2\, (2\pi)^4\delta^{4}(P+K-Q)
     \, D^0_{1\leftrightarrow2}\,\bm{B}^{H}_s\cdot \bm{B}^{H}_t\\ \nonumber
  &=& \frac{g_{3}^{2}}{96\pi^{3}}  \frac{T^{5}}{v_s^{6}}
    \sum_{s,t\neq 0} b^H_s b^H_t
    \int \dif \Gamma_{xyz}\, |{\cal M}_3(x,y,z)|^2\,
      (2\pi)^4\delta^{4}(X+Y-Z) 
      \, D^0_{1\leftrightarrow2}\, \bm{B}^{H}_s\cdot \bm{B}^{H}_t\\ \nonumber 
  &=& \frac{g_{3}^{2}}{96\pi^{3}}\frac{T^{5}}{v_s^{6}}
    \sum_{s,t\neq 0} b^H_s b^H_t M^H_{st} \, , 
\end{eqnarray}
where $\dif \Gamma_{pkq}$ is the 3-particle Lorentz-invariant phase
space, $X = (x_0,{\bm x}) = (p_0/T,v_s{\bm p}/T), etc.$, 
$D^0_{1\leftrightarrow2} = f_0(x)f_0(y)(1+f_0(z))$ and
$\bm{B}^{H}_t=\big[B^H_{t}(x)\bm{x} + B^H_t(y)\bm{y} -
B^H_t(z)\bm{z}\big]/\sqrt{3}$. As before, we can define
\begin{equation}
D_H = (A_{11}^H)^2 (M^{H}_{st})^{-1}_{11}\, , 
\end{equation}
which allows us to write the thermal conductivity as
\begin{equation}
 \kappa_H \geq\frac{\langle \bm{g}|\varepsilon_p \bm{v}_p
   \rangle^{2}}{\langle \bm{g}|{\hat C}|\bm{g} \rangle}
   = \frac{128}{3\pi}\frac{\gamma^{2}}{g_{3}^{2}}
       \frac{T^{2}}{v_s^{2}}\, D_H
   =  \frac{256\sqrt{2}}{25\pi^{3}\xi^{2}}\, \kappa_0
     \Big(\frac{T}{T_{F}}\Big)^{2} \, D_H\left(T/T_F\right)\, , 
\end{equation}
where $\xi\simeq 0.44$ is defined in Section \ref{sec:exc}. $D_H$ is 
a dimensionless function of $T/T_F$ which has to be calculated 
numerically. The convergence of $D_H$ with the number of terms in 
the trial function is shown in Fig.~\ref{kappa_bose}. We observe that 
the result in converging, but not nearly as fast as it did in the case
of fermionic quasi-particles. The temperature dependence of $D_H$ is 
very weak and $\kappa_H\sim T^2$. In Fig.~\ref{kappa_both} we combine
our results for the thermal conductivity of the dilute Fermi gas 
at unitarity in the high and low temperature limits. The dashed
line indicates the location of the phase transition. We note that
the low and high temperature results do not match very well 
near $T_c$. This suggests that additional processes or excitations
must become relevant in this regime. We also note that at $T_c$
critical fluctuations are expected to lead to a divergence in 
the thermal conductivity \cite{Hohenberg:1977ym}.

\section{Shear Viscosity and the Prandtl Number}
\label{sec:prandtl}

\subsection{Shear Viscosity}

 In order to compare the magnitude of thermal and momentum diffusion 
in dilute atomics gases we also need to determine the viscosity of 
the gas. The shear viscosity of the dilute Fermi gas at unitarity
has been calculated in both the high temperature \cite{Bruun:2005}
and low temperature limits \cite{Rupak:2007vp}. Here, we quickly 
summarize the results and present a simple numerical formula that 
can be used to compute the contribution from binary scattering 
of atoms for all values of $k_Fa$ and $T/T_F$. 
 
 The viscous correction to the stress tensor is 
\begin{equation}
\delta T_{ij} =  \nu\int \dif \Gamma v_p \frac{p_i p_j}{p} \delta f_p
= - \eta V_{ij}\, , 
\end{equation}
where $V_{ij} = \partial_i V_j + \partial_j V_i - 2/3\delta_{ij} 
\nabla \cdot V$ and $V_i$ is the velocity of the fluid. We use
an ansatz for $\delta f_p$ analogous to the one that appeared 
in the calculation of the thermal conductivity, 
\begin{equation}
\delta f_p = -\frac{f_0(p)(1\pm f_0(p))}{T} \chi_{ij} V_{ij}\, . 
\end{equation}
The shear viscosity $\eta$ is then given by 
\begin{equation}
\eta  = \frac{\nu}{5 T} \int \dif \Gamma f_0(p)(1\pm f_0(p)) 
  \frac{v_p}{p}  p_i p_j \chi_{ij} \equiv \frac{3 T^2}{5} 
  \langle \tilde{p}_{ij}| \chi_{ij}\rangle\, , 
\end{equation}
where $\tilde{p}_{ij} = v_p(p_i p_j - \frac{1}{3} \delta_{ij} p^2)/p$ 
and we have used the inner product defined previously. The linearized 
Boltzmann equation for a fluid perturbed by a small shear stress is 
given by 
\begin{eqnarray}
\frac{\dif f}{\dif t} &=& 
  \frac{f_0(p)(1\pm f_0(p))}{2T} |\tilde{p}_{ij}\rangle V_{ij}\\ \nonumber
  &=& C[f_p] \equiv \frac{f_0(p)(1\pm f_0(p))}{2T} 
                     \hat C|\chi_{ij}\rangle V_{ij} \, .
\end{eqnarray}
As before, we can combine these results into a form that is suitable 
for a variational treatment. We write
\begin{equation}
 \eta = \frac{3 T^2}{5}\frac{\langle\chi_{ij}|\tilde{p}_{ij}\rangle^2}
                         {\langle\chi_{ij}|\hat{C}|\chi_{ij}\rangle}\, . 
\end{equation}
and $\chi_{ij} = \frac{p_{ij}}{\Delta^2}\tilde{g}(x)$. We expand 
$\tilde{g}(x) = \sum_{s=0}c_s C_s(x^2)$ where the polynomials $C_s(x^2)$
are defined in analogy with \Eqn{ortho} but with the weight factor 
$x^4$ replaced by $x^6$, see Appendix \ref{app:trial}. In the 
case of binary atomic scattering the variational bound is 
\begin{equation}
\label{eta_F}
\eta_F \geq \frac{2^{7/2}\eta_{0}}{45}
            \bigg(\frac{T}{T_{F}}\bigg)^{3/2}
            D^{\eta}(k_Fa,T/T_F)\, , 
\end{equation}
where $\eta_0 = (mT_F)^{3/2}$ and $D^{\eta}(k_Fa,T/T_F)=(A_0^{\eta})^{2}
(M^{\eta})^{-1}_{00}$. The normalization integral is 
\begin{equation}
\label{A_00_eta}
    A_0^{\eta}=\int\dif x f_0(x)(1-f_0(x)) x^6
              =  z \int \dif x\ x^6 \frac{e^{x^2}}{(e^{x^2}+z)^2}\, , 
\end{equation}
where $z = e^{\mu/T}$ is the fugacity and the collision matrix 
elements $M_{st}^\eta$ are given by 
\begin{equation}
\label{M_st_eta}
    M^{\eta}_{st}=\int \frac{\dif^{3}x\dif^{3}y}{(2\pi)^{6}}\,\, \dif \Omega
  \frac{mTa^{2}y}{1 + 2mTa^{2} y^2}
  \, D^0_{2\leftrightarrow2}\, \widehat{C}_{s}\cdot \widehat{C}_{t}
\end{equation}
with $D^0_{2\leftrightarrow2} = f_0(x_1)f_0(x_2)(1-f_0(x_3))(1-f_0(x_4))$ and
\begin{equation}
   \widehat{C}_{t}=\big[C_{t}(x_{1})x_{1}^{ij}+C_{t}(x_{2})x_{2}^{ij}
   -C_{t}(x_{3})x_{3}^{ij}-C_{t}(x_{4})x_{4}^{ij}\big]/2 \, . 
\end{equation}
As in the case of thermal conductivity we can analytically compute 
$D^\eta$ by restricting the calculation to the lowest order orthogonal 
polynomial, and by considering the high temperature limit. We find
\begin{equation}
\eta_{F}=
\begin{cases}
    \frac{5\sqrt{mT}}{32a^{2}\sqrt{\pi}} & \ \ \ |a|\rightarrow0 \\
    \frac{15(mT)^{3/2}}{32\sqrt{\pi}}    & \ \ \ |a|\rightarrow\infty\ .
\end{cases}
\label{eta_highT}
\end{equation}
In the superfluid phase the shear viscosity is dominated by binary 
phonon scattering. The leading low temperature behavior in the 
unitarity limit was calculated in \cite{Rupak:2007vp}. Their result 
is 
\begin{equation}
\eta_{H}= 5.06 \times 10^{-6} \xi^{7/2}
\left(\frac{T_F}{T}\right)^5\eta_{0} \, . 
\end{equation}

\subsection{Prandtl Number}

\begin{figure}[t]
\begin{center}
\includegraphics[width=8.8cm]{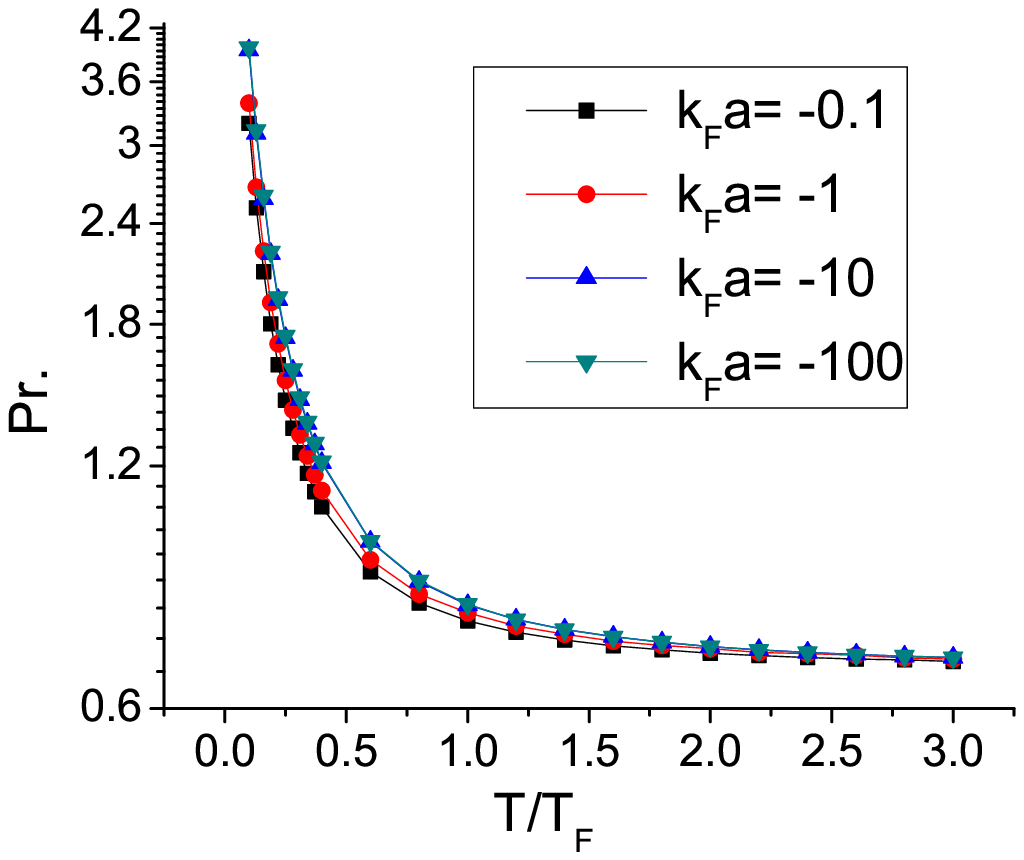}
\includegraphics[width=8.8cm]{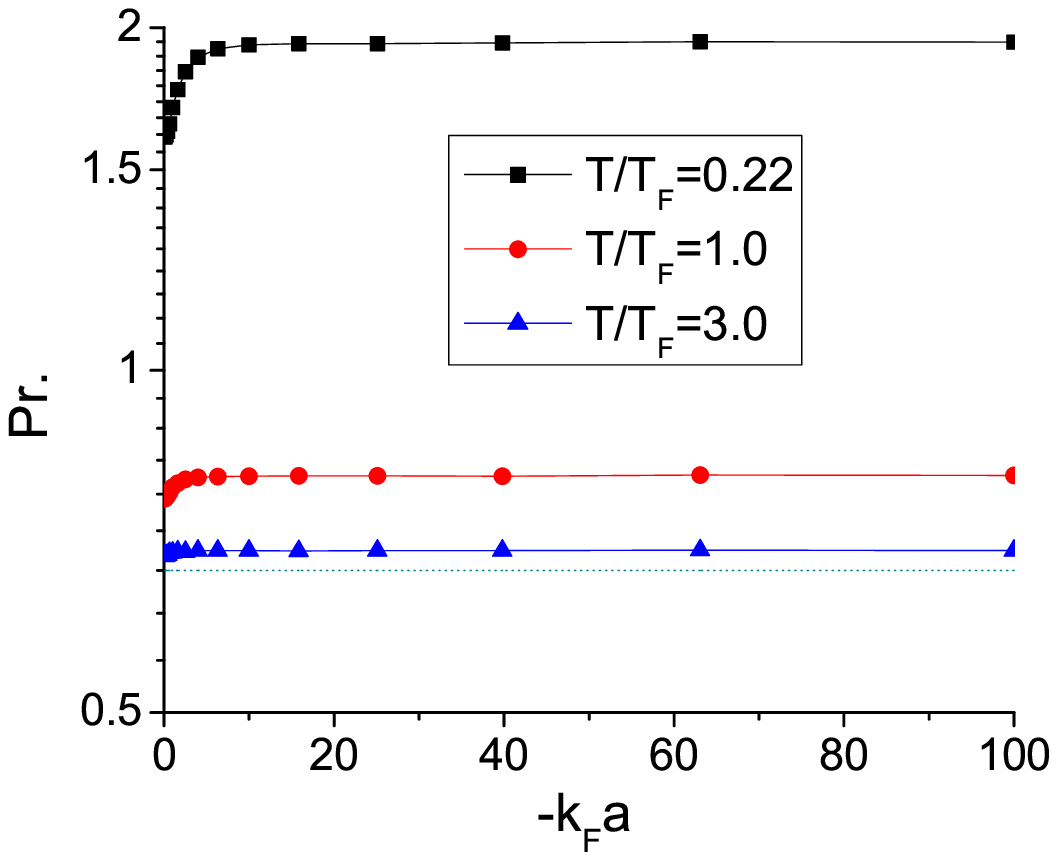}
\end{center}
\caption{The left panel shows the Prandtl number as a function of 
$T/T_{F}$ for different values of $k_Fa$. The right panel shows
the Prandtl number as a function of $k_F a$ for several values 
of $T/T_F$. The dashed line shows the high temperature limit 
${\it Pr}=2/3$. 
\label{pr1}}
\end{figure}

 The Prandtl number ${\it Pr}=c_{P}\eta /(\rho\,\kappa)$ characterizes 
the relative importance of thermal and momentum diffusion. Here, $\rho$
is the mass density of the fluid and $c_P$ is the specific heat at 
constant pressure. In the high temperature limit the specific heat
can be calculated using the Virial expansion, and at low temperature 
it can be computed using the effective phonon lagrangian. The results 
are summarized in Appendix \ref{app:vir}. The leading term at high 
temperature is determined by the equation of state of an ideal gas
of spin up and down atoms. In this limit we get  $c_p = 5n/2$ and 
$\rho = m n$. The high temperature limits of $\kappa$ and $\eta$ are
given in \Eqn{kappa_highT} and \Eqn{eta_highT}. Putting these results
together we get
\begin{equation}
{\it Pr} = 2/3 \qquad \qquad T\gg T_F \,
\end{equation}
independent of the size of the scattering length. This result
is remarkable because it implies that the Prandtl ratio agrees
with the classical gas result even though the shear viscosity
and thermal conductivity are not classical. This result can be
traced to the fact that in the large $T$ limit the $x$ and
$y$ integrals in \Eqn{collintN} and \eqn{M_st_eta} factorize. The interaction 
only depends on $y$ (related to the momentum transfer) and the leading
$y$ dependence of ${\bm B}_1\cdot {\bm B}_1$ and $\hat{C}_0
\cdot \hat{C}_0$ is identical. This implies that both $\eta$
and $\kappa$ can be expressed in terms of the same transport
cross section, and that there is no dependence on the
scattering length in the Prandtl ratio.  In Fig.~\ref{pr1}
we plot the Prandtl number as a function of $T/T_F$ for negative values 
of $k_F a$.  As discussed in more detail in Appendix~\ref{app:vir}, the 
Virial expansion is smooth across the BCS/BEC transition, but our 
approximation for the bound state energy of the atom-atom dimer,
$E_B=1/(2ma^2)$, breaks down as we go to far towards the BEC side
of the transition. For this reason we only show the Prandtl number
for negative values of $k_Fa$. We observe that the Prandtl number 
increases significantly as the temperature is lowered, and that 
it is only weakly dependent on $k_Fa$. 

\section{Sound Attenuation}
\label{sec:sound_att}

\begin{figure}[t]
\begin{center}
\includegraphics[width=10cm]{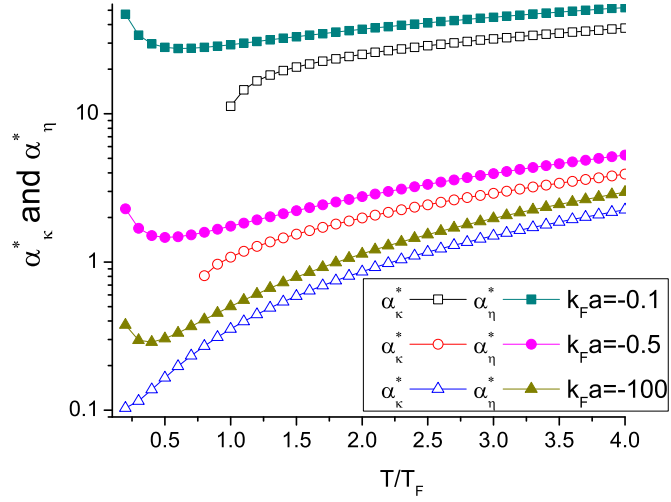}
\caption{$\alpha_\eta^*$ and $\alpha_\kappa^*$ as a function of $T/T_F$.
Note that the ratio of the two approaches a constant ($3/4$) as we
go to larger values of $T/T_F$.
\label{fig:att}}
\end{center}
\end{figure}

  Using the results for the shear viscosity and the thermal 
conductivity we can also compute the sound attenuation coefficient. 
The intensity of a plane wave decreases as $e^{-2\gamma x}$ where 
$\gamma=\alpha\omega^2$ is the absorption coefficient and $\omega$
is the frequency of the wave. In the normal phase the parameter 
$\alpha$ is given by \cite{Landau:fl}
\begin{equation}
\label{att_1}
   \alpha =\frac{\gamma}{\omega^{2}}=\frac{1}{2\rho c^{3}}
    \left[ \frac{4}{3}\eta+\zeta + 
   \rho\,\kappa\left(\frac{1}{c_{V}}-\frac{1}{c_{P}}\right)\right]\, , 
\end{equation}
where $\rho$ is the mass density of fluid, $c$ is the speed of sound 
and $\zeta$ is the bulk viscosity. At unitarity $\zeta = 0$ and we can
write 
\begin{equation}
\alpha = \frac{2 \eta}{3 \rho c^3}
  \left[1 + \frac{3}{4}\frac{\Delta c_P}{Pr}\right]
  \equiv \frac{\eta_0}{2\rho c^3}(\alpha_\eta^* + \alpha_\kappa^*)
\end{equation}
where $\Delta c_P=(c_P-c_V)/c_V$ and $\eta_0=(mT_F)^{3/2}$ as in 
\Eqn{eta_F}. The quantities $\alpha_\eta^*$ and $\alpha_\kappa^*$ 
are plotted in Fig.~\ref{fig:att}.

In the high temperature limit the Prandtl number is ${\it Pr} \sim 2/3$ 
and $\Delta c_P \sim 2/3$. This implies that that the contribution to the 
sound attenuation coefficient due to thermal conductivity is $3/4$ of
the contribution due to shear viscosity.  Also, both ${\it Pr}$ and 
$\Delta c_P$ are only weakly dependent on the scattering length, see 
Fig.~\ref{fig:att}. As the temperature is decreased the sound absorption
coefficient is increasingly dominated by shear viscosity.

 Below $T_c$ there are two sound modes, ordinary (first) sound and
second sound. The absorption coefficient for first sound is analogous
to \Eqn{att_1}, except that the contribution from $\kappa$ is absent. 
This implies that damping of first sound is entirely due to shear 
viscosity. The absorption coefficient of second sound is given by 
\cite{Wilks:1966,Putterman:1974}
\begin{equation}
\label{att_2}
\alpha_2 = \frac{1}{2 \rho c_2^3}\frac{\rho_s}{\rho_n}
 \left[\frac{4}{3}\eta + \zeta_2 - \rho(\zeta_1 + \zeta_4) 
   + \rho^2 \zeta_3 + \rho\frac{\rho_n}{\rho_s}\frac{\kappa}{c_P}\right]
\end{equation}
where $\rho_{n,s}$ are the normal and superfluid densities, $c_2$ is the 
speed of second sound and $\zeta_i$ are the four different coefficients 
of bulk viscosity in the superfluid. Onsager's symmetry principle implies
$\zeta_4=\zeta_1$, and conformal symmetry requires that at unitarity 
$\zeta_1=\zeta_2=0$. The remaining coefficient, $\zeta_3$, vanishes
as $\zeta_3\sim T^3$ as the temperature goes to zero \cite{Escobedo:2009bh}. 
Using $\eta \sim 1/T^5$ and $\kappa \sim T^2$ we conclude that the 
damping of second sound is also dominated by shear viscosity.

We should note that at low temperature the phonon mean free
path becomes very long, and the normal fluid ceases to behave
hydrodynamically. A rough estimate of the relaxation times for
shear viscosity and thermal conductivity is $\tau_\eta\simeq
\eta/(nT)$ and $\tau_\kappa \simeq m\kappa/(c_PT)$. This implies
that the relaxation time for shear viscosity grows very quickly,
$\tau_\eta\sim 1/T^6$, whereas the relaxation time for thermal
conductivity grows more slowly, $\tau_\kappa\sim 1/T^2$.

\section{Conclusions}
\label{sec:concl}

 In this work we studied the thermal conductivity of a dilute Fermi 
gas. Our main results are summarized in Fig.~(\ref{kappa_both}). We
observe that the thermal conductivity scales as $\kappa \sim m^{1/2}T^{3/2}$ 
in the high temperature limit and as $\kappa\sim m^{1/2}T_F^{3/2}(T/T_F)^2$ 
in the low temperature limit. The Prandtl ratio equals $2/3$ in the high 
temperature limit, and increases as the temperature is lowered. This, 
together with the fact that $\Delta c_P=(c_P-c_V)/c_V$ decreases at 
low temperature, implies that the sound attenuation length in the 
strongly coupled normal fluid regime $T_c<T<T_F$ is dominated by shear 
viscosity. In order to assess the feasibility of measuring the sound 
attenuation length we consider the ratio of wavelength $\lambda$ 
to the attenuation length $\gamma^{-1}$. We find 
\begin{equation}
\label{lam_gam}
\lambda\gamma = \frac{4\pi}{\sqrt{3\xi}}
  \Big(\frac{k}{k_F}\Big)\Big(\frac{\eta}{n}\Big)
 \left[ 1 + \frac{3}{4}\frac{\Delta c_P}{{\it Pr}}\right]
 \simeq 11.4  \Big(\frac{k}{k_F}\Big)\Big(\frac{\eta}{n}\Big)\, ,
\end{equation}
where we have used $\xi\simeq 0.44$ and $\Delta c_P/{\it Pr}
\simeq 0$. A rough estimate of the minimum value of $\eta/n$ 
can be obtained by extrapolating the kinetic theory result 
down to $T=T_c$. We get 
\begin{equation}
 \Big(\frac{\eta}{n}\Big)_{\it min}
\simeq \frac{45\pi^{3/2}}{64\sqrt{2}}
  \left(\frac{T_c}{T_F}\right)^{3/2} \simeq 0.15\, . 
\end{equation}
Finally, we can express $k/k_F$ in terms of the wavelength 
and the trap parameters. We obtain
\begin{equation}
\frac{k}{k_F} = \frac{R_z}{\lambda} 
  \frac{\pi\beta^{2/3}}{(3N)^{1/3}} \, ,
\end{equation}
where $R_z$ is the longitudinal cloud size, $\beta=R_\perp/R_z$
is the trap asymmetry, and $N$ is the number of particles. For 
the experiment reported in \cite{Joseph:2006} we get $k/k_F\simeq 
0.046$ and $\lambda\gamma\simeq 0.08$. This implies that a
measurement of the sound attenuation length is within reach.
Higher sensitivity can be achieved by using more elongated 
traps or smaller wavelengths. 

 There are a variety of improvements that need to be made
once experimental data become available. The most important
is to replace the plane wave approximation which is the basis
of \Eqn{lam_gam} by a more detailed calculation of the wave
profile in a given trap geometry. It will also be interesting 
to study attenuation of first and second sound in the 
superfluid phase. 

Acknowledgments: This work was supported in parts by the US
Department of Energy grant DE-FG02-03ER41260. We would like 
to thank C.~Manuel, M.~Mannarelli and J.~Thomas for useful 
discussions.

\begin{appendix}
\section{Trial Functions}
\label{app:trial}

 In this appendix we provide explicit expressions for some of the 
orthogonal polynomials and associated normalization constants. The 
orthogonality condition for the polynomials $B^{(F,H)}_s(x^2)$ is
(see \Eqn{ortho})
\begin{equation}
A^{(F,H)}_{s} \delta_{st} =  \int \dif x\, 
   f_0(x)(1\pm f_0(x)) x^4 B^{(F,H)}_s(x^2) B^{(F,H)}_t(x^2)\, , 
\end{equation}
where $B^{(F,H)}_0(x^2)=1$ and 
\begin{equation}
B^{(F,H)}_s(x^2) = x^{2s} + \sum_{i=0}^{s-1} a^{(F,H)}_{si} x^{2i}
\end{equation}
for $s\geq 1$. We will determine the coefficients $a^{(F,H)}_{si}$
iteratively. We define
\begin{equation}
c^{(F,H)}_n = \int \dif x\, f_0(x)(1\pm f_0(x)) x^{4+n}\, , 
\end{equation}
so that $A_0 = c_0$. We also find
\begin{eqnarray}
a_{10} &=& -c_2/c_0\, ,  \\ \nonumber
a_{20} &=& (c_2 c_6 - c_4^2)/(c_0 c_4 - c_2^2)\, ,  \\ \nonumber
a_{21} &=& (c_0 c_6 - c_2 c_4)/(c_0 c_4 - c_2^2)\, , \\ \nonumber
\vdots &&
\end{eqnarray}
and $A_1 = c_4 - c_2^2/c_0$. The integrals $c_n$ depend on the 
quasi-particle dispersion relation. In the case of the phonons with
$E_p = v p$, we can evaluate $A_1$ to be
\begin{equation}
A^H_1 = \frac{4\pi^4}{15}\ .
\end{equation}
In the case of the fermion with $E_p = p^2/2 m$ and a chemical potential,
we can only evaluate $A_1$ in the case of large temperature which gives
\begin{equation}
A^F_1 \simeq z \frac{15 \sqrt{\pi}}{16}
\end{equation}
where $z = e^{\mu/T}$ is the fugacity.

The orthogonal condition of the shear viscosity's trial function polynomials 
$\tilde{g}(x) = \sum c_s C_s(x^2)$ is different from the one of thermal 
conductivity. Here, the polynomial is starting from $s=0$ term since this 
set of polynomials satisfies the constraint of momentum 
conservation law automatically (due to the tensor nature of the 
kernel for shear viscosity).  The orthogonality condition is then
\begin{equation}\label{shear_ortho}
    A^{\eta}_{s}\delta_{st}=\int\dif x
    f_0(x)(1-f_0(x))x^{6}C_{s}(x^{2})C_{t}(x^{2})
\end{equation}
where the trial function $C_{s}(x^2)=x^{2s}+\sum_{i=0}^{s-1} a_{si} x^{2i}$. 
By the orthogonality condition~\Eqn{shear_ortho} and the same recursion method 
that we used before, it is easy to get the coefficients $a_{si}$ and also
$A_0^\eta$ which is given by
\begin{equation}
A_0^\eta = \int\dif x  f_0(x)(1-f_0(x))x^6 = z \int \dif x\ 
x^6 \frac{e^{x^2}}{(e^{x^2}+z)^2} \simeq z \frac{15 \sqrt{\pi}}{16}
\end{equation}
where the analytic result is valid in the high temperature limit.  Note that
$A_1^F = A_0^\eta$ in the high temperature limit; however, this is a 
mathematical coincidence coming from the similarity of the different
integrals and the properties of the $\Gamma$-function.

\section{Thermodynamic Properties}
\label{app:vir}

\subsection{Virial Expansion}

In the high temperature limit, $T\gg T_{F}$, the pressure can 
be expanded in powers of the fugacity $z=\me^{\mu/T}$. This 
expansion is known as the Virial expansion. We write 
\begin{equation}
    P(\mu,T)=\frac{\nu T}{\lambda^{3}}\sum_{l=1}^{\infty}b_{l}z^{l}\, , 
\end{equation}
where $\lambda = \sqrt{2\pi/mT}$ is thermal wave length an $\nu$ is 
the spin degeneracy factor. The first Virial coefficient is $b_{1}\equiv 1$.
The second Virial coefficient was first calculated by Beth and Uhlenbeck 
\cite{Beth:1936}, see also \cite{Ho:2004,Lee:2005is}. Here we use 
these results in order to compute $\Delta c_P$. There are two contributions 
to $b_2$. The first is due to quantum statistics, which becomes important 
as $n\lambda^{3}\sim 1$, the other is related to the atomic interaction. 
The correction due to quantum statistics is
\begin{eqnarray}
  b_{l}^{0} =\begin{cases}
  l^{-\frac{5}{2}} &  \ \ \ \ \ \ \mathrm{Bosons}\, ,  \\
 (-1)^{l+1}l^{-\frac{5}{2}}& \ \ \ \ \ \ \mathrm{Fermions}\, . 
 \end{cases}
\end{eqnarray}
The interacting contribution to the second Virial coefficient is 
\begin{equation}
\label{b2_int}
    b_{2}-b^{0}_{2}=
 \sqrt{2}\sum_{n}(\me^{-\epsilon_{n}/T}-\me^{-\epsilon_{n}^{0}/T})\, , 
\end{equation}
where $\epsilon_{n}$ is the energy of the $n$'th two-particle state
in the center of mass frame and $\epsilon_{n}^{0}$ is the corresponding 
non-interacting energy. In the universal regime these energies are 
determined by the scattering length only. We get \cite{Lee:2005is} 
\begin{equation}
b_2 - b^0_2 = 
\left\lbrace \begin{array}{cl}
    \frac{1}{\sqrt{2}}\me^{x^2}[1-{\rm erf}(|x|)] & x<0\, ,  \\ \nonumber
    \sqrt{2}e^{x^2} - \frac{1}{\sqrt{2}}\me^{x^2}[1-{\rm erf}(x)]\; & x>0\, ,
   \end{array}\right.
\end{equation}
where $\rm{erf}(x)$ is the error function, $x = \lambda/(a\sqrt{2\pi})$ 
and the bound state energy in the $x>0$ case was assumed to be 
$E_B = 1/(ma^2)$.  We can write down the asymptotic forms in the limits 
of zero and infinite scattering lengths as
\begin{equation}\label{}
    b_{2}-b^{0}_{2}=\begin{cases}
         -\frac{a}{\lambda} & \ \ \ a\rightarrow 0^- \, ,\\
         \sqrt{2}\,e^{1/(mTa^2)} & \ \ \ a\rightarrow 0^+ \, , \\
         \frac{1}{\sqrt{2}}(1+\frac{\sqrt{2}}{\pi}\frac{\lambda}{|a|})  
     & \ \ \ a\rightarrow\pm\infty\, . 
         \end{cases}
\end{equation}
The interaction part of $b_{2}$ approaches a finite value as $a\rightarrow
\infty$, and is of the same order as the effects of quantum statistics. One 
can also see that the total $b_2$ will start off negative for small and 
negative scattering length, but eventually becomes positive as $a\rightarrow 
-\infty$. This implies that there is a value of the scattering length for 
which the effects of the interaction and of quantum statistics cancel. 
For positive scattering lengths, the contribution of the interaction is 
positive and always larger than the quantum correction, so that $b_2>0$ 
for all $a>0$. In the BEC limit ($a \rightarrow 0^+$) the binding energy 
of the bound state is large and $b_2$ becomes large also. In this limit, 
however, universality breaks down and the binding energy cannot 
be expressed in terms of the scattering length only.

\begin{figure}
\begin{center}
\includegraphics[width=8.8cm]{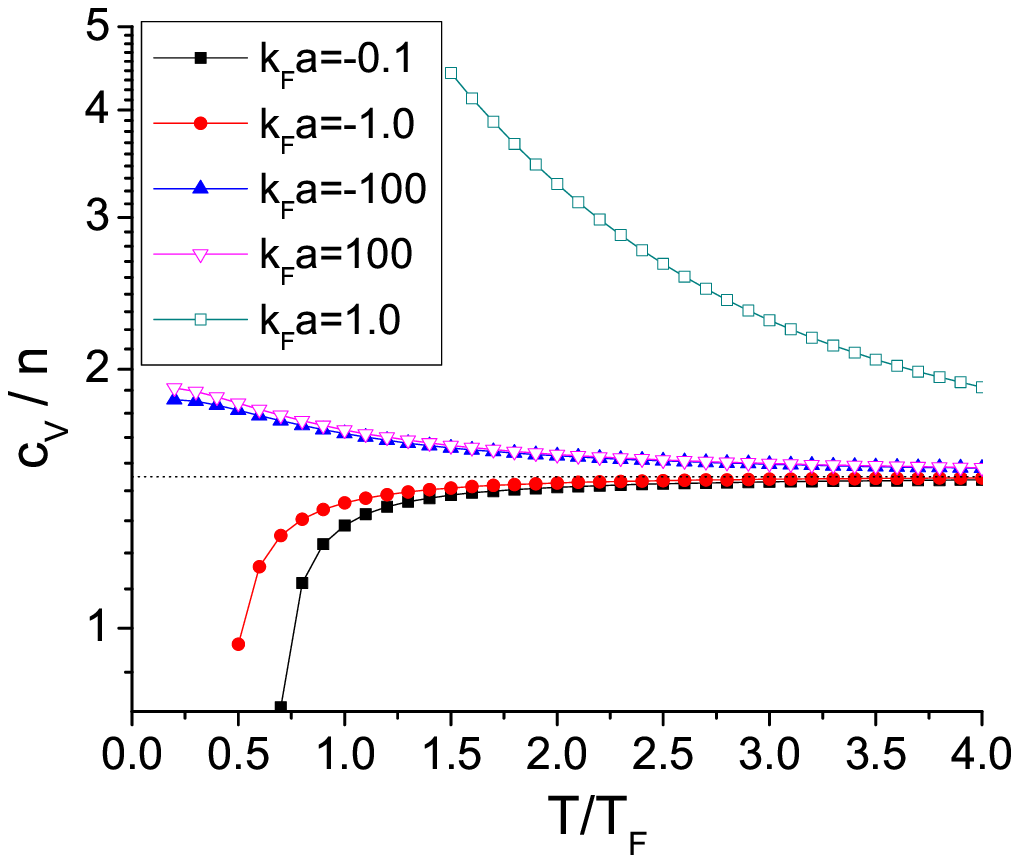}
\includegraphics[width=8.8cm]{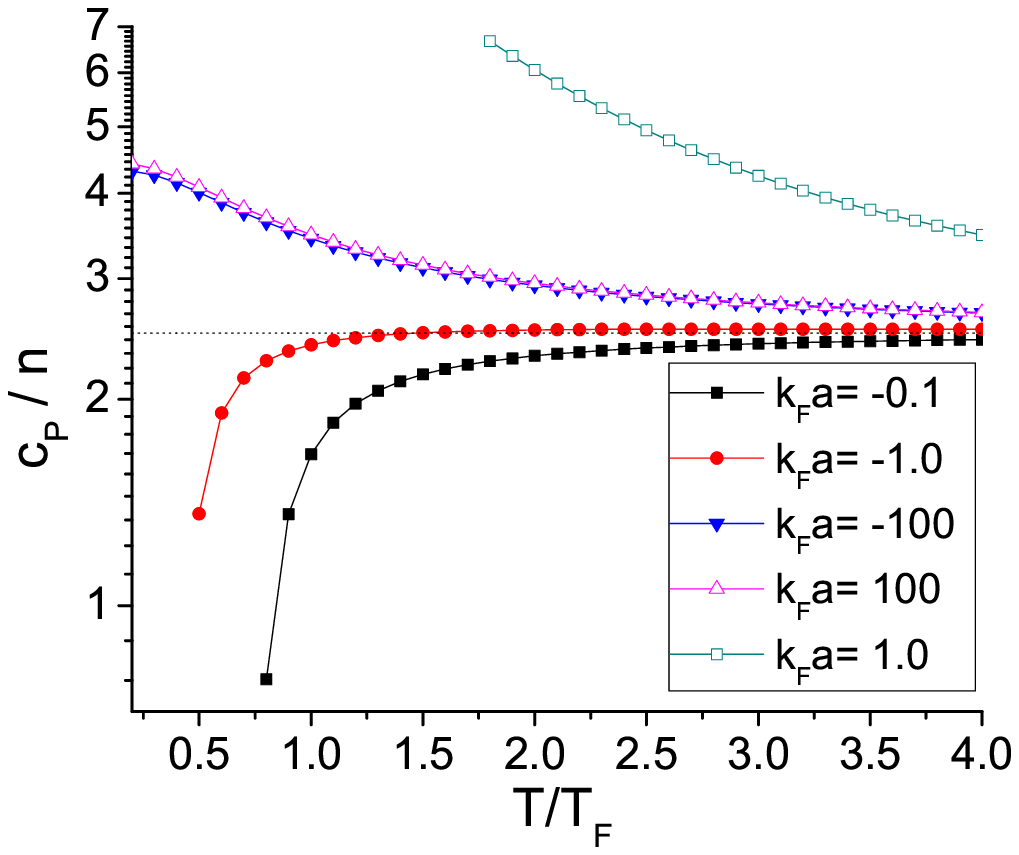}
\end{center}
\caption{\label{fig:cvcp}
The left panel shows $c_{V}$ in units of $n_0$ as a 
function of $T/T_F$ $n_{0}$. Here, $n_0=\nu z/\lambda^3$ is the 
density of the free gas. The right panel shows the temperature 
dependence of $c_P$.}
\end{figure}

\subsection{Specific Heat}

 The specific heat can be expressed in terms of partial derivatives 
of the pressure. The independent thermodynamic variables are $(\mu,T)$.
The first derivatives with respect to theses variable determine the
entropy density and pressure
\begin{equation}
\label{s_and_n}
s=\left.\frac{\partial P}{\partial T}\right|_{\mu}\, , \hspace{1cm}
n=\left.\frac{\partial P}{\partial \mu}\right|_{T}\, . 
\end{equation}
In order to compute the specific heat at constant volume we use 
$V=N/n$ and write 
\begin{equation}
\label{c_v_part}
 c_{V}= T\left.\frac{\partial s}{\partial T}\right|_{V}
      = \frac{\partial(s,V)}{\partial(T,V)}
      = \frac{\partial(s,V)/\partial(T,\mu)}{\partial(T,V)/\partial(T,\mu)}
      = T\left[\left.\frac{\partial s}{\partial T}\right|_{\mu}
          -\frac{[(\partial n/\partial T)|_{\mu}]^2}
                {(\partial n/\partial \mu)|_{T}}\right]\, , 
\end{equation}
where the Jacobian is defined as 
\begin{equation}
\frac{\partial(a,b)}{\partial(c,d)} = {\rm Det}
   \left(\begin{array}{cc}
       \partial a/\partial c\big|_d & \partial a/\partial d\big|_c \\ \nonumber
       \partial b/\partial c\big|_d & \partial b/\partial d\big|_c
       \end{array}\right)\, . 
\end{equation}
In order to compute $c_P$ we make use of the relation between 
$c_P-c_V$ and the thermal expansion coefficient $(1/V)(\partial V/
\partial T)|_P$. This relation is given by 
\begin{equation}
\label{cP}
   c_{P}-c_{V}=-T\frac{[(\partial V/\partial T)|_{P}]^{2}}
                      {(\partial V/\partial P)|_{T}}\, . 
\end{equation}
The partial derivatives are 
\begin{equation}
  \left. \frac{\partial V}{\partial T}\right|_{P}
   =\frac{s}{n}\left.\frac{\partial n}{\partial \mu}\right|_{T}
      -\left.\frac{\partial n}{\partial T}\right|_{\mu} \, , 
\hspace{1cm}
   \left. \frac{\partial V}{\partial P}\right|_{T}
      =\left. \frac{\partial n}{\partial \mu}\right|_{T}\, , 
\end{equation}
which gives
\begin{equation}\label{}
    c_{P}=c_{V}+T\frac{\Big[
    \frac{s}{n} (\partial n/\partial \mu)\big|_{T}
     -(\partial n/\partial T)\big|_{\mu}\Big]^{2}}
              {(\partial n/\partial \mu)\big|_{T}}\, . 
\end{equation}

\subsubsection{High Temperature Phase}

 In the high temperature phase we use the Virial expansion of the 
pressure including terms up to second order, $P=\frac{\nu T}{\lambda^{3}}
(z+b_{2}z^2)$. Using the results of the previous section we find 
\begin{eqnarray}
  c_{V} &=& \frac{\nu z}{\lambda^{3}}\left[\frac{3}{2}
      + \frac{15}{4}zb_{2}-zT\frac{\partial b_{2}}{\partial T}
      + zT^{2}\frac{\partial^{2} b_{2}}{\partial^{2} T}\right]\, ,  
      \\ \nonumber
  c_{P} &=& c_{V}+\frac{\nu z}{\lambda^{3}}\left[1+5zb_{2}
            - 2zT\frac{\partial b_{2}}{\partial T}\right]\, . 
\end{eqnarray}
We recover the result for a classical non-interacting gas by setting 
$b_{2}=0$ and $n=\frac{\nu z}{\lambda^{3}}$. We show the result for 
an interacting gas in Fig.~\ref{fig:cvcp}. We observe that the 
results are smooth across the BCS/BEC transition. We note, however, 
that on the BEC side the Virial expansion breaks down at
temperatures larger than the Fermi temperature. This is related 
to the presence of a deeply bound state.

\subsubsection{Superfluid Phase}

 At zero temperature the pressure is given by
\begin{equation}
P  = \frac{4\sqrt{2}}{15\pi^2 \xi^{3/2}} m^{3/2}\mu^{5/2}\, ,
\end{equation}
and the density is 
\begin{equation}
n =\left.\frac{\partial P}{\partial \mu}\right|_{T}
  =\frac{2\sqrt{2}m^{3/2}}{3\pi^{2}\xi^{3/2}}\mu^{3/2}\, . 
\end{equation}
The leading low-temperature correction arises from phonons. The 
phonon contribution to the pressure is 
\begin{equation}\label{}
P_{H}=\frac{\pi^{2} T^{4}}{90}\Big(\frac{3m}{2\mu}\Big)^{3/2}
\end{equation}
Using these results we find 
\begin{eqnarray}
  c_{V} &=& \frac{2\pi^2 T^{3}}{15} \Big(\frac{3m}{2\mu}\Big)^{3/2} 
                   +\mathcal{O}(T^{7}) \, , \\\nonumber
  c_{P} &=& c_{V}+\mathcal{O}(T^{7})\, . 
\end{eqnarray}
These results imply that $\Delta c_p$ vanishes at low temperature
much faster than $c_P$ and $c_V$. 
\end{appendix}

\end{document}